\documentclass[12pt,a4paper]{article}

\usepackage[english]{babel}

\usepackage[letterpaper,top=2cm,bottom=2cm,left=3cm,right=3cm,marginparwidth=1.75cm]{geometry}

\usepackage{amsmath}
\usepackage{graphicx}
\usepackage[colorlinks=true, allcolors=blue]{hyperref}

\usepackage{graphicx}
\usepackage{subcaption}

\usepackage{tabularx} 
\usepackage{array}

\usepackage{multirow} 
\usepackage{booktabs} 
\usepackage{makecell} 
\usepackage{array} 
\usepackage{multirow} 

\usepackage{ragged2e} 

\usepackage{caption}
\usepackage{array} 
\usepackage{cellspace} 
\usepackage{array} 
\usepackage{cellspace} 
\usepackage{colortbl} 
\definecolor{lightgray}{gray}{0.8} 

\title{Hybrid Wi-Fi/PDR Indoor Localization with Fingerprint Matching}
\author{CHUNYI ZHANG\footnote{Author’s email: zhangchunyi529@gmail.com}, ZONGWEI LI, XIAOQI LI}

\date{}

\begin{document}
\maketitle

\begin{abstract}
Indoor position technology has become one of the research highlights in the Internet of Things (IoT), but there is still a lack of universal, low-cost, and high-precision solutions. This paper conducts research on indoor position technology based on location fingerprints and proposes a practical hybrid indoor positioning system. In this experiment, the location fingerprint database is established by using RSS signal in the offline stage, the location algorithm is improved and innovated in the online stage. The weighted k-nearest neighbor algorithm is used for location fingerprint matching and pedestrian dead reckoning technology is used for trajectory tracking. This paper designs and implements an indoor position system that performs the functions of data collection, positioning, and position tracking. Through the test, it is found that it can meet the requirements of indoor positioning.
\end{abstract}


\section{Introduction}
With the advancement of wireless communication technology and the popularity of smartphones, location-based service (LBS) has attracted much attention \cite{kenan2021comparative, wu2025atomicity, xiao2025parallelizing}. LBS refers to software services that use geographic data and information to provide services or information to users, typically used in navigation software, social networking services, location-based advertising, tracking systems, mobile commerce, and personalized weather services \cite{2, aguilarSmartContractFamilies2024}. Its implementation is based on determining the user's location information, namely positioning technology. Traditional positioning technologies use satellites for positioning, such as the Global Positioning System (GPS) and China's Beidou Satellite Navigation System \cite{lu2021brief, arceriSoundConstructionEVM2024}. Satellite-based positioning technology is mature, accurate, stable, and has a good positioning effect in the outdoor environment. However, because of the low power of the satellite signal, it is difficult to penetrate the walls of the building. Coupled with its own scattering, reflection, multipath effect, etc., the signal has been severely attenuated in the indoor environment \cite{4}. Due to the influence of the complex indoor environment, the positioning effect has been far from meeting the needs of people. In addition, indoor positioning requires higher positioning accuracy, so the development of indoor positioning systems has become a research hotspot in the Internet of Things.

With the wider application prospects, the demand for indoor positioning is also constantly increasing \cite{farahsari2022survey}. In daily life, navigation services at railway stations, airports, large shopping centers, and libraries all require positioning technology; in public safety and emergency response, timely and accurate positioning of people can strive for valuable time for rescue operations; in social needs, the rise of mobile Internet has also driven the demand for location information on social platforms, such as WeChat's function of sharing real-time location; in social care, visually impaired people will also benefit from indoor positioning systems. In terms of future applications, indoor positioning systems can be used for mobile computing, big data analysis, robot navigation applications, etc.

In light of the significant demand for indoor LBS, numerous companies and enterprises both domestically and internationally have carried out relevant research and proposed their own solutions. In 2011, Google added an indoor navigation function to its mobile phone map application, using information such as Wi-Fi and cellular base stations. In 2013, Apple developed its own indoor maps using a wide range of iPhone users based on low-power Bluetooth technology. In 2019, Apple incorporated an Ultra Wide Band (UWB) chip into the new iPhone 11 mobile phone, released to support UWB high precision indoor positioning \cite{coppens2022overview}. At the same time, commercial giants like Baidu, Tencent, and Alibaba in China are also competing in this field \cite{ding2021smartloc}. During the 12th Five-Year Plan period, China launched the "Xihe" programme, with the aim of building a system that can provide indoor 3m and outdoor 1m precision positioning services. During the 13th Five-Year Plan period, China initiated a number of indoor positioning projects within the field of earth observation \cite{baoguo2023research}. 

Despite the rapid development and variety of indoor position technology, traditional solutions still have persistent challenges \cite{20}:
(1) Accuracy issues. It should be noted that low-cost technologies, such as Wi-Fi fingerprints, are subject to large measurement errors due to signal fluctuations.
(2) Cost issues. UWB and other high-precision solutions require dedicated hardware and the deployment of the relevant base station equipment, which is expensive.
(3) Real-time issues. Complex algorithms such as particle filters take a long time to calculate, and the response delay of mobile terminals is notable. Furthermore, the high energy consumption limits its practical value.

In response to existing issues of indoor positioning technology, this study proposes a hybrid positioning method that integrates Wi-Fi fingerprinting and pedestrian dead reckoning (PDR). Firstly, it introduces the classification of indoor positioning technology and analyzes the current research status of location fingerprinting technology. Secondly, the collection of location fingerprints has been meticulously designed, and the dynamic weighted WKNN algorithm has been refined to optimize the fingerprint matching strategy. In combination with the PDR algorithm, low-latency trajectory updates are achieved using mobile phone sensor data. Then, a positioning solution based on Android is implemented, which verifies the practicality of the solution with the existing Wi-Fi infrastructure, providing a cost-effective solution for indoor positioning. Finally, it summarizes the strengths and weaknesses of the proposed scheme, along with the direction of subsequent research improvement.

The main contributions of this study are:
\begin{itemize}
\item \textbf{Improved positioning algorithm:} It proposes a dynamic weighting strategy based on the reciprocal of Euclidean distance to optimize the positioning accuracy of the traditional KNN algorithm.
\item \textbf{Real-time positioning and tracking through multi-sensor fusion:} It Integrates Wi-Fi fingerprint positioning with the PDR algorithm, combined with mobile phone accelerometer and direction sensor data, to achieve real-time position updates and trajectory tracking.
\item \textbf{Low-cost Android system implementation:} It develops client applications based on the Android platform, using the existing Wi-Fi infrastructure without the need for additional hardware deployment, and builds a lightweight server using the Spring Boot framework.
\item \textbf{Complete experimental verification in public settings:} It conducts system testing in actual indoor environments and provides detailed experimental data and results.
\end{itemize}
This study aims to provide valuable references for the research of indoor positioning technology and reproducible solutions for the optimization of subsequent positioning algorithms.

\section{Related Work}
The main indoor positioning techniques can be divided into three categories: computer vision-based, sensor-based, and communication-based positioning \cite{10, ayubSoundAnalysisMigration2024}. Computer vision-based systems employ omnidirectional cameras, 3D cameras, or built-in smartphone cameras to extract information about the indoor environment. They then use various image processing algorithms for feature extraction and matching \cite{caiEnablingCompleteAtomicity2024}. Although this method boasts high accuracy, it also has significant financial implications and is susceptible to external factors that can disrupt signals. Sensor-based methods principally refer to Pedestrian Dead Reckoning (PDR), which uses data from accelerometers, gyroscopes, magnetometers, etc., and can estimate the user's position based on past positions \cite{chenDemystifyingInvariantEffectiveness2024}. However, this method is frequently integrated with other positioning techniques due to the accumulation of drift-related errors over time and the requirement for initial calibration \cite{wang2024ContractsentryStaticAnalysis}.

A plethora of communication-based indoor positioning techniques are available, including Radio Frequency Identification (RFID), Wi-Fi, Bluetooth, visible light communication (VLC), and ultra-wide band (UWB) \cite{11}. RFID has a limited operational range and offers limited system versatility \cite{shang2022overview}. Wi-Fi based systems mostly use location fingerprinting, which uses existing infrastructure and is suitable for implementation. However, it requires regular updating of the fingerprint library. In terms of Bluetooth positioning techniques, they are similar to Wi-Fi positioning \cite{jiang2021fingerprint}, but they have weak noise immunity and poor positioning stability. VLC-based systems use LED or fluorescent lamps that are already in place in the building. This approach is cost-effective but relies on established communication facilities for location information \cite{13}. UWB-based systems have the ability to achieve centimeter-level accuracy; unfortunately, the equipment is costly and the penetration rate is low \cite{qi2024current}.

Of the technologies mentioned above, the location fingerprint has been shown to be an effective technique due to its simplicity and ease of deployment. In 2000, P Bah et al. from Microsoft Research designed the Radar system based on Received Signal Strength (RSS) signals \cite{bahl2000radar}. They used the K-Nearest Neighbor (KNN) algorithm to estimate the target's location, thereby establishing the framework for indoor location fingerprinting algorithms. Some scholars have also investigated the feasibility of other types of signal, including the signal-to-noise ratio \cite{wallbaum2005benchmarking}, signal impulse response information \cite{jin2010indoor}, and channel state information \cite{wang2016csi}. In addition to the use of a single type of signal, fused RSS fingerprint localization using additional sensors or devices to aid positioning has also been thoroughly researched. For example, Dr Ban et al. combined PDR, magnetic field, and Wi-Fi fingerprints to estimate location by comparing pedestrian sensor and fingerprint values through particle filters, which greatly improved localization accuracy \cite{18}. Tian et al. have developed a probabilistic mathematical model to address the issue of high volatility in RSS signals \cite{wen2015fundamental}. This model uses the volume of the RSS signal on the surface of a convex packet to construct a location fingerprint. They have also proposed a model for the propagation of RSS signals \cite{tian2017improve}. In order to establish a fingerprint database, Y. Mo et al. proposed the use of the event-based fingerprint clustering method \cite{mo2012occurrence}. This method involves the clustering of fingerprint information collected from the reference node on multiple occasions. The result of this is an improvement in the accuracy of fingerprint location. In the field of online stage matching, novel algorithms and theories are constantly emerging. For instance, Carlos et al. have proposed a method that utilizes a priori information in conjunction with the Support Vector Machine (SVM) algorithm \cite{figuera2012advanced}. Experimental results have demonstrated that this approach produces superior positioning accuracy compared to the KNN algorithm.

\section{Methodology}
\begin{figure}
\centering
\includegraphics[width=0.99\linewidth]{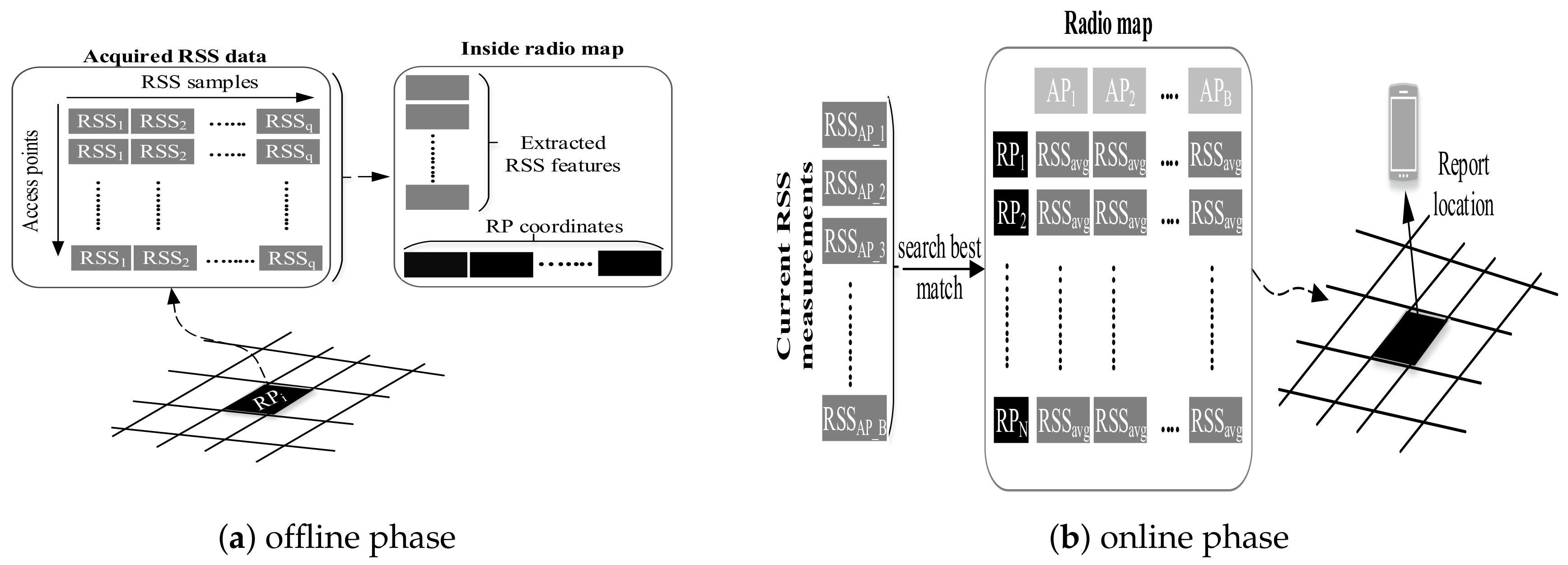}
\caption{\label{fig:stage}Schematic diagram of location fingerprint positioning stage}
\end{figure}

Location fingerprint technology is an implementation scheme of indoor positioning technology based on Wi-Fi \cite{sun2025FIRESmartContract}. Despite the advantages of Wi-Fi, including its extensive deployment, ease of promotion and low cost, its use as a communication signal is generally limited to a single antenna with a small bandwidth, which hinders its compatibility with traditional positioning and ranging methods. The advent of location fingerprint technology has effectively addressed this issue \cite{ezhumalai2021efficient, grossmanPracticalVerificationSmart2024}. Location fingerprinting is defined as the process of associating a location with a unique identifier, known as a "fingerprint". The location is defined as the coordinate of the positioning point, and the fingerprint is defined as the feature information of the positioning point. As demonstrated in Figure \ref{fig:stage}, the process of location fingerprint localization is comprised of two distinct stages: offline training and online localization \cite{16, hanOSwapPreservingAtomicity2026}. In the offline phase, location fingerprints are collected and a wireless map is established (see Section \ref{sec:subsection1}). In the online phase, fingerprint samples are best matched with fingerprints in the wireless map to obtain location coordinates (see Section \ref{sec:subsection2}). Furthermore, the system integrates a PDR algorithm to achieve the trajectory tracking function (see Section \ref{sec:subsection3}), based on the initial position obtained through the positioning function. Finally, Section \ref{sec:subsection4} provides a comprehensive overview of the system architecture.

\subsection{Location Fingerprint Database Construction}
\label{sec:subsection1}
In principle, any information feature that can identify a location can be used to create a location fingerprint \cite{heCodeNotNatural2024}. Such features include the multipath structure of the Wi-Fi signal at a given location, the monitoring of an access station or a public mobile communication base station at a location, the RSS from the base station that can be monitored at a location, the round-trip time of the signals during communication at a location, and the delay \cite{22, huangAdvancingWeb302024}, etc.  The location fingerprint employed in this system is the RSS (Radio Signal Strength) signal, which is also one of the most commonly used signal features. RSS is a special term for RF (Radio Frequency) signals, which refers to the power output of a transmitter received by a reference antenna at a certain distance from the transmitting antenna \cite{23, jiaoSurveyEthereumSmart2024}. The strength of the received signal is used to determine the distance between the point of the signal and the point of reception. The calculation formula for RSS is shown in Equation 1 \cite{24}:
\begin{align}
    \mathrm{x}=10 \log _{10} \frac{P}{1 m W}
\end{align}

The unit of measurement that corresponds to RSS is decibel milliwatt (dBm). This is employed to represent power level in decibels (dB), with reference to one milliwatt (mW). The converted RSS typically exhibits a negative tendency, ranging from near 0 dBm (excellent signal) to less than -100 dBm (poor signal). However, its attenuation is not simply a linear relationship with an increase in distance. RSS is intrinsic to Wi-Fi devices and is easily obtainable. It is characterised by low power and low cost, which renders it ideal for Wi-Fi indoor positioning.

The establishment of a location fingerprint database is the process of creating a wireless map \cite{feng2024review}. The initial step involves the division of the indoor space into a grid of m rows and n columns, with a conventional grid size of 1 m. Multiple APs are then positioned within the designated area, such as Wi-Fi routers and mobile phones with hotspots enabled. The coordinates and the corresponding fingerprint data for each grid point are then meticulously recorded. The fingerprint data in this context pertains to the RSS signal strengths from various Access Points (APs) received at a specific location. The RSSs from multiple APs collectively constitute the location fingerprint of the location. In the context of the acquisition process, the instability of Wi-Fi signals and the influence of the surrounding environment render the AP data collected at the same location likely to vary. Consequently, multiple repeated measurements are required, the mean is taken, and finally the radio map is constructed.

\subsection{Localization Implemented by WKNN}
\label{sec:subsection2}
The commonly used location fingerprint localisation algorithms are divided into two categories: deterministic and probabilistic localisation algorithms \cite{26}. Deterministic localisation algorithms are predicated on the comparison of statistical values of the fingerprint sample vectors with a location fingerprint database in order to estimate the location. In contrast, probabilistic localisation algorithms achieve localisation by calculating the probability distribution of the fingerprint sample vectors on the radio map. In addition to the two aforementioned types of algorithms, there are studies that employ neural network algorithms for indoor positioning. These studies utilise a similar scenario analysis to generate localisation estimates.

In deterministic fingerprint localisation, the closest match is taken to be the estimated position coordinate by comparing the measured RSS sample with the RSS stored in the database. This is the simplest deterministic localisation algorithm, which is essentially the Nearest Neighbour (NN) algorithm. The NN algorithm is employed to identify the nearest neighbour to the sample fingerprint in the radio map, and subsequently estimates its corresponding location coordinates as the location of the mobile device. The primary objective of the algorithm under consideration is to ascertain the distance between the sample to be tested and all the location fingerprints in the radio map. The calculation of this distance can be performed in a variety of ways, including the Manhattan distance, the Cosine similarity, the Jaccard similarity and the Euclidean distance. The calculation of distance using the Euclidean distance formula is outlined in Equation 2 \cite{28}.
\begin{align}
    D_{i T}=\left(\frac{\sum_{i=1}^{n}\left(\left|R S S_{T}-R S S_{i}\right|\right)^{2}}{n}\right)^{\frac{1}{2}}, i=1,2,3, \ldots, n
\end{align}

$RSS_T$ is defined as the RSS strength of the sample testing point, $RSS_i$ is the RSS strength of each reference point, and $n$ is the number of reference points. The NN algorithm is characterised by simplicity and ease of implementation; however, it exhibits suboptimal positioning accuracy. The issue can be resolved through the utilisation of KNN (K-Nearest Neighbour) and WKNN (Weighted K-Nearest Neighbour) algorithms.

NN, KNN and WKNN algorithms are classified as supervised learning algorithms for classification and regression. They are considered to be "inert" learning algorithms, as they do not involve an explicit learning process. It is imperative that the spatial distance of RSS between the existing reference points in the radio map and the sample test points is calculated by all parties. The NN algorithm selects the coordinates of the reference point corresponding to the smallest distance as the result output. The KNN algorithm selects k reference points with the smallest distance and uses the average of the coordinates of these k points as the result output. The calculation formula of the KNN algorithm is shown in Equation 3.
\begin{align}
    \left(x_{T}, y_{T}\right)=\frac{\sum_{i=1}^{K}\left(x_{i}, y_{i}\right)}{K}
\end{align}

Where $(x_{T}, y_{T})$ is the coordinates of the sample measurement point, and $(x_{i}, y_{i})$ refers to the coordinates of the $i$-th reference point. The WKNN algorithm also selects k points, but assigns weights to these points and uses the weighted average of the coordinates as the output. In consideration of the impact of distance on RSS strength, this experiment employs the reciprocal of the Euclidean distance as a weighting metric. The formula for calculating the weight is outlined in Equation 4.
\begin{align}
    w_{i T}=\frac{\frac{1}{D_{i T}}}{\sum_{i=1}^{K} \frac{1}{D_{i T}}}
\end{align}

Where $w_{i T}$ refers to the weight of the $i$-th reference point. The calculation of the weighted average of the coordinates of these reference points is then used to determine the coordinates of the sample measurement point. The calculation formula is as outlined in Equation 5.
\begin{align}
    \left(x_{T}, y_{T}\right)=\sum_{i=1}^{K} w_{i T}\left(x_{i}, y_{i}\right)
\end{align}

\begin{figure}
    \centering
    \begin{subfigure}[b]{0.48\textwidth}
        \includegraphics[width=\textwidth]{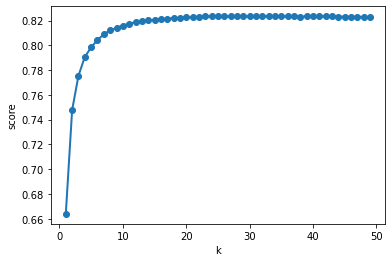}
        \caption{\label{fig:Comparison chart of K values}Comparison chart of K values}
    \end{subfigure}
    \hfill
    \begin{subfigure}[b]{0.48\textwidth}
        \includegraphics[width=\textwidth]{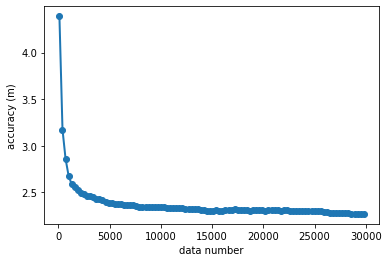}
        \caption{\label{fig:Comparison chart of data volume}Comparison chart of data volume}
    \end{subfigure}
    \hfill
    \caption{\label{fig:parameter}Parameter Comparison}
\end{figure}

The accuracy of the algorithm's positioning is also related to the choice of K value and the size of the data volume. In order to establish the parameters of the algorithm with greater precision, the WKNN algorithm is selected for subsequent verification in the simulation environment. Figure \ref{fig:Comparison chart of K values} shows the relationship between the K-value and the cross-validation score. It has been demonstrated that an increase in the K-value is associated with an enhancement in the cross-validation score. When the K-value exceeds 5, the score attains a value greater than 0.8, and consequently, the subsequent system selects K-value 5. Figure \ref{fig:Comparison chart of data volume} illustrates the correlation between the quantity of data and the cross-validation score. It is evident that an augmentation in the data set results in an enhancement in the system's accuracy.

\subsection{Trajectory Tracking Implemented by PDR}
\label{sec:subsection3}
The pedestrian dead reckoning algorithm has its origins in the sailing process. Sailors must master their own orientation, and they use a compass and other tools to record the direction and speed of their vessel. At the same time, they use the correction of the sign building to judge their position. When the PDR algorithm is extended to indoor positioning technology, it mainly refers to the use of inertial measurement units to detect the acceleration, angular velocity, magnetic force, pressure, and other information of the target in motion, and estimate the direction and step size based on this information, so as to achieve the tracking of the target position \cite{jiang2022smartphone}.

\begin{figure}
\centering
\includegraphics[width=0.58\linewidth]{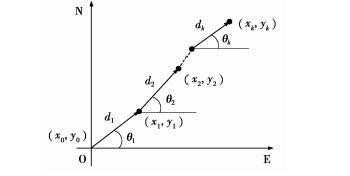}
\caption{\label{fig:Schematic diagram of PDR algorithm}Schematic diagram of PDR algorithm}
\end{figure}
As demonstrated in Figure \ref{fig:Schematic diagram of PDR algorithm}, the fundamental principle of the PDR algorithm is to assess the pedestrian's step frequency based on the period of acceleration and to estimate the step length through the utilisation of a pertinent model. The angle information is then obtained from the direction sensor, and the initial position, the azimuth angle $\alpha $, and the step length d are employed to determine the position of the subsequent step. This is subsequently integrated with the information regarding the number of steps to facilitate the continuous calculation of the current position. The calculation formula is shown in equation 6 \cite{wen2023enhanced}.
\begin{align}
    \left\{\begin{array}{l}
x_{n}=x_{n-1}+d \times \cos \alpha \\
y_{n}=y_{n-1}+d \times \sin \alpha
\end{array}, n>1\right.
\end{align}

The PDR algorithm is comprised of three primary components: gait detection, step length projection, and orientation projection. The basis of gait detection is the periodicity of the acceleration trajectory of the user during walking, with the threshold being detected by the modal values of the three axes of the acceleration sensor, thus deriving the corresponding gait characteristics. There are two models for step length projection: the static model and the dynamic model. The system utilises the static model, which establishes a constant step size. The orientation projection requires the rotation angle, which can be obtained directly by the direction sensor of the mobile device. The WKNN algorithm is employed to obtain the initial position coordinates, and the PDR algorithm is then used to calculate the next coordinate. The positioning trajectory is obtained by continuously updating the UI to annotate the next coordinate on the user interface.

\subsection{System Architecture}
\label{sec:subsection4}
\begin{figure}
\centering
\includegraphics[width=0.68\linewidth]{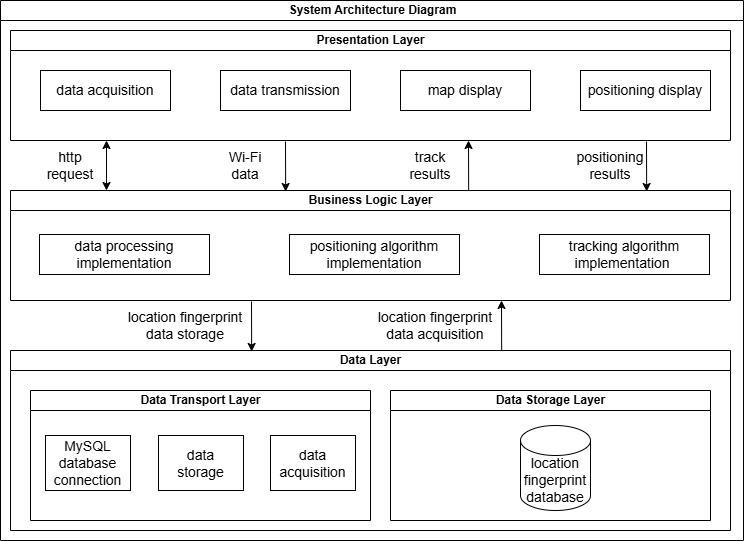}
\caption{\label{fig:System Architecture Diagram}System Architecture Diagram}
\end{figure}

The system necessitates the deployment of an APP client, tasked with the collection of location fingerprint data, the implementation of algorithms for online positioning by the backend server, and the establishment of a database for the storage of location fingerprint data. In order to facilitate data transmission and interaction between the application programming interface (API) client and the backend server, the backend server is required to add and manage data in the database. The three-tier architecture (see Figure \ref{fig:System Architecture Diagram}) under discussion is designed with the following layers: representation, business logic and data.

The representation layer is situated at the uppermost level, and its function is to provide the user with an interface for interactive operations. The primary functions of the representation layer are: data acquisition, data transmission, map display and positioning display. The data acquisition  function includes the coordinate setting of the acquisition point, Wi-Fi scanning and target AP matching. The data transmission function comprises the transmission of Wi-Fi signal strength information and coordinate information to the server, and the reception of positioning coordinate data. The map display function pertains to the display of the initial map drawn by the custom view. The positioning display function includes displaying positioning coordinates on the map, displaying positioning coordinates in the form of Toast pop-up window, and displaying positioning track on the map.

The business logic layer constitutes the core of the system structure, with the function of making logical judgements and executing operations on specific problems. The business logic layer is the conduit between the representation layer and the data layer. Subsequent to receiving user commands and data from the representation layer, it connects to the data layer and performs logical processing and algorithmic operations on the received data, thereby realising the positioning function. The principal functions of the business logic layer comprise data processing, implementation of the positioning algorithm, and implementation of the positioning tracking algorithm. The functions of data processing include the reception of data information from the client, the parsing of said data information, and the transmission of the positioning coordinates. The positioning algorithm refers to the calling of the WKNN algorithm to obtain the positioning coordinates, and the positioning tracking algorithm refers to the calling of the PDR algorithm to obtain the real-time moving coordinates.

The data layer is divided into two parts: the data transfer layer and the data storage layer. The function of the data layer is to store and read system data. The primary functions of the data transfer layer are as follows: the establishment of a connection with the MySQL database, the storage of data, and the acquisition of data. The data storage layer is responsible for the fixed storage of data, including the location fingerprint database.

The system utilises a MySQL database for the purpose of data storage. The data to be stored comprises the location coordinates (X and Y) and the signal strength of the Wi-Fi deployment point \cite{singh2021machine, kumarVulnerabilitiesSmartContracts2024}. The data types are of the float variety, and none of them permit null values. The design is illustrated in Table \ref{tab:Data Table Design}.

\begin{table}
    \centering
    \caption{\label{tab:Data Table Design}Data Table Design}
    \begin{tabular}{llll}
        \toprule
        \textbf{Item} & \textbf{Data Type} & \textbf{Not NULL} & \textbf{Description} \\
        \midrule
        X & float & yes & X coordinate \\
        Y & float & yes & Y coordinate \\
        AP1 & float & yes & Signal strength of AP1 \\
        AP2 & float & yes & Signal strength of AP2 \\
        AP3 & float & yes & Signal strength of AP3 \\
        AP4 & float & yes & Signal strength of AP4 \\
        AP5 & float & yes & Signal strength of AP5 \\
        \bottomrule
    \end{tabular}
\end{table}

\section{Experiments \& Results}
\subsection{Simulation Environment Comparison}
\begin{figure}
\centering
\includegraphics[width=0.58\linewidth]{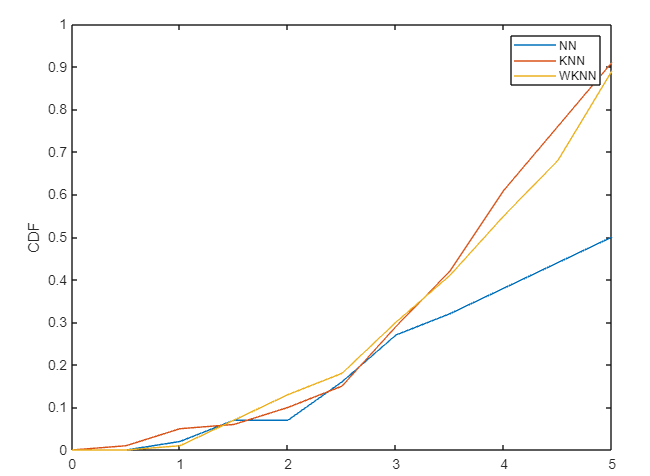}
\caption{\label{fig:CDF comparison chart}CDF comparison chart}
\end{figure}

A simulation platform was built based on Matlab, and 1000 samples were used for location prediction with NN, KNN, and WKNN algorithms, respectively. The RSS measurements were generated by the logarithmic path loss model, the AP number was set to 4, and the Euclidean distance was used for weighting. The cumulative distribution function (CDF) is shown in Figure \ref{fig:CDF comparison chart}. The CDF is defined as the integral of the probability density function, thus providing a comprehensive description of the probability distribution of a real random variable X. It is important to note that the higher the probability of identifying a region, the higher the positioning accuracy. The CDF curve of the WKNN algorithm demonstrates superiority over those of the NN and KNN algorithms.

\begin{figure}
\centering
\includegraphics[width=0.58\linewidth]{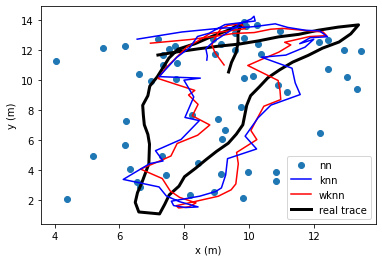}
\caption{\label{fig:Trajectory comparison chart}Trajectory comparison chart}
\end{figure}

Figure \ref{fig:Trajectory comparison chart} gives the real paths in the experimental environment and the trajectories predicted by the NN, KNN and WKNN algorithms. The blue dots represent the localisation points predicted by the NN algorithm, the blue trajectories represent the paths predicted by the KNN algorithm, and the red trajectories represent the paths predicted by the WKNN algorithm. After comparison, it can be concluded that the KNN and WKNN algorithms fit better than the NN algorithm.

\subsection{Field Environment Testing}
The field test of this study was conducted in the eastern sector of the ground floor of the eighth building of China University of Mining and Technology (Beijing), as illustrated in Figure 5.1. The experimental area was meticulously delineated into 180 localisation points, employing a 1m×1m grid configuration. In the test environment, there are more than eight AP points, and five of these with the strongest signal are selected as the transmitter. The receiver is a Huawei Mate10 cell phone with Android version 10, running the Android client for data collection. The server and database have been deployed on a Windows 10 laptop, and it is imperative that both the server and client utilise the same LAN. The database management system selected for this study was MySQL, and a selection of the collected data is presented in Table \ref{tab:Partial Data Collection}.
\begin{table}
    \centering
    \caption{\label{tab:Partial Data Collection}Partial Data Collection}
    \begin{tabular}{ccccccc}
        \toprule
        \textbf{X} & \textbf{Y} & \textbf{AP1} & \textbf{AP2} & \textbf{AP3} & \textbf{AP4} & \textbf{AP5} \\
        \midrule
        0 & 0 & -46 & -41 & -55 & -68 & -67 \\
        1 & 0 & -53 & -54 & -59 & -55 & -69 \\
        1 & 0 & -56 & -51 & -67 & -63 & -69 \\
        2 & 0 & -67 & -52 & -72 & -63 & -48 \\
        2 & 0 & -72 & -73 & -92 & -38 & -53 \\
        3 & 0 & -72 & -62 & -86 & -44 & -46 \\
        3 & 0 & -37 & -52 & -47 & -77 & -78 \\
        4 & 0 & -37 & -52 & -47 & -77 & -78 \\
        2 & 1 & -37 & -42 & -42 & -65 & -82 \\
        2 & 1 & -37 & -42 & -42 & -65 & -82 \\
        2 & 2 & -41 & -40 & -56 & -62 & -75 \\
        2 & 2 & -42 & -30 & -48 & -66 & -73 \\
        1 & 2 & -43 & -31 & -49 & -64 & -73 \\
        1 & 2 & -36 & -33 & -47 & -67 & -77 \\
        0 & 3 & -37 & -28 & -44 & -59 & -68 \\
        \bottomrule
    \end{tabular}
\end{table}

\begin{figure}
    \centering
    \begin{subfigure}[b]{0.3\textwidth}
        \includegraphics[width=\textwidth]{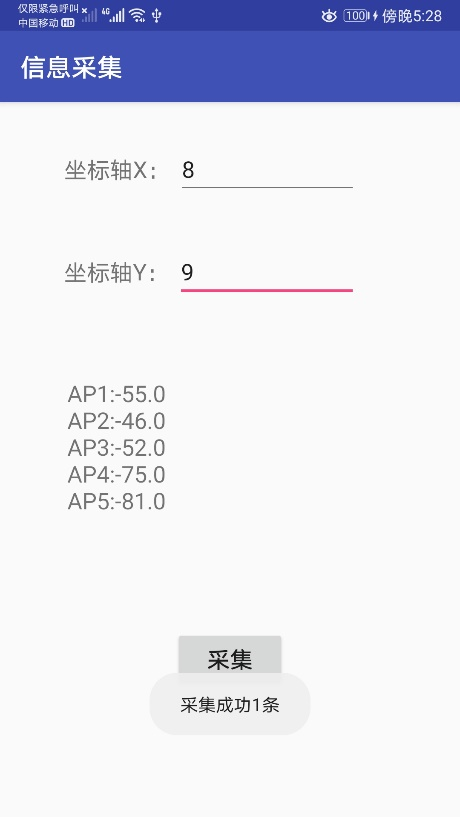}
        \caption{\label{fig:Data Collection}Data Collection}
        \label{fig:sub1}
    \end{subfigure}
    \hfill
    \begin{subfigure}[b]{0.3\textwidth}
        \includegraphics[width=\textwidth]{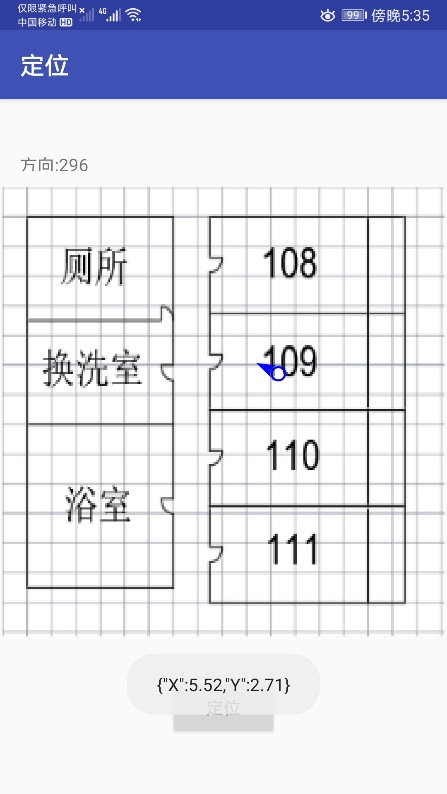}
        \caption{\label{fig:Localisation}Localisation}
        \label{fig:sub2}
    \end{subfigure}
    \hfill
    \begin{subfigure}[b]{0.3\textwidth}
        \includegraphics[width=\textwidth]{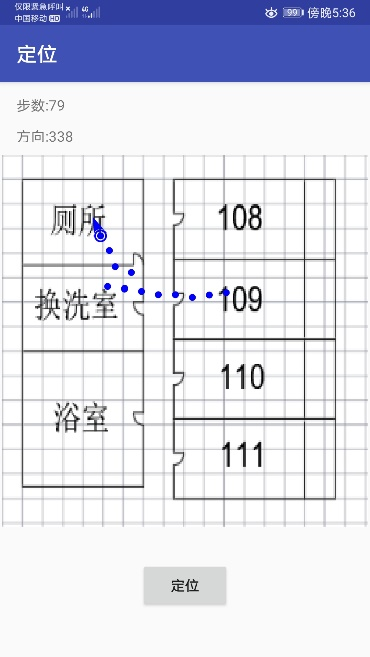}
        \caption{\label{fig:Trajectory Tracking}Trajectory Tracking}
        \label{fig:sub3}
    \end{subfigure}
    \caption{\label{fig:Functional Testing}Functional Testing}
    \label{fig:test}
\end{figure}

Following the testing phase, it was established that the data collection, localisation and trajectory tracking functions are all operational and respond within a brief timeframe. When data collection is conducted at the designated positioning point, as delineated on the map prepared previously, the precise coordinates of the positioning point must be entered into the system. Thereafter, the system will collect the signal strength data from the five AP points and store it (see Figure \ref{fig:Data Collection}). Subsequent to the collection of all data from the positioning points, the positioning test is conducted at the designated test point. This process successfully displays the coordinates of the positioning point and its position on the map (see Figure \ref{fig:Localisation}). Subsequent to the acquisition of the initial position, the handheld mobile device moves at a constant speed to obtain the number of steps and direction of movement, thereby successfully realising trajectory tracking (see Figure \ref{fig:Trajectory Tracking}). The mean response time for data acquisition is 20 seconds, with the variability of this metric being attributable to the Wi-Fi signal strength of the AP deployment point. The mean response time for localization is 3 seconds, and the mean response time for trajectory tracking is 1 second.

\section{Conclusion}
This paper preliminarily explores the classification of indoor position technology and the implementation scheme based on location fingerprint technology. Common indoor localization techniques include computer vision-based localization, sensor-based localization, and localization based on different communication technologies such as RFID, Wi-Fi, Bluetooth, visible light communication, and ultra-wideband. Location fingerprinting has emerged as a prominent technique in this field, using widely deployed Wi-Fi networks. This study proposes a pragmatic hybrid indoor positioning system that integrates Wi-Fi fingerprinting and pedestrian dead reckoning, providing a cost-effective solution that optimizes accuracy, real-time performance, and deployment complexity \cite{liASTRODetectingAccess2025, liDemoEnhancingSmart2024}. By using existing Wi-Fi infrastructure and smartphone sensors, the system can perform data acquisition, localization, and trajectory tracking functions without additional hardware \cite{priftiSmartContractVulnerability2024, suDiSCoDecompilingEVM2025}. The proposed dynamic weighted WKNN algorithm enhances traditional fingerprint matching by incorporating distance-based weighting, while sensor fusion with PDR effectively achieves real-time trajectory tracking \cite{wangContractCheckCheckingEthereum2024}. The experimental results obtained from field tests have confirmed the feasibility of the system, with stable localization accuracy and low latency.

However, there are still some issues in this system that can be optimized and improved. The data acquisition process requires manual collection, which is both cumbersome and susceptible to constant maintenance requirements. The PDR algorithm generates a cumulative error, which increases over time and requires updating of the localization points. Furthermore, the system's functionality is restricted to the collection and localization in the 2D plane, thereby lacking the capacity to support 3D localization. Therefore, future research should focus on the following:
\begin{itemize}
\item \textbf{Explore the realization of automatic acquisition and updating of the fingerprint database.}  The integration of models such as deep learning is imperative in addressing the deficiencies of the prevailing system, which is characterized by manual data acquisition and the cumbersome maintenance of static fingerprint databases. The use of real-time WiFi signals during user engagement will facilitate the automated generation and dynamic updating of the fingerprint database.
\item \textbf{Explore the realization of PDR error compensation.} The experimental investigation will be conducted to explore the establishment of a feedback mechanism that has the capacity to suppress the cumulative error of the PDR algorithm. The objective of this investigation is to enhance the accuracy and stability of long-time trajectory tracking by using timed or fixed-point calibration.
\item \textbf{Explore the realization of 3D localization.} Existing Wi-Fi fingerprinting and PDR algorithms focus on horizontal movement and lack the capacity to model data in the vertical direction. Combining more information such as barometer, altimeter, or geomagnetic fingerprints, future research will explore the possibility of realizing 3D localization and enhance the system's practicality in various application scenarios.
\end{itemize}

Through these efforts, this study aims to provide a more comprehensive theoretical guide and implementation scheme for indoor localization.

\section*{Acknowledgments}
The author would like to express sincere gratitude to Prof. Wu Yaqin for her invaluable guidance and support during the foundational stages of this research.

\bibliographystyle{plain}
\bibliography{bibliography}

\end{document}